\documentclass[review,times,number,12pt]{elsarticle}

\usepackage{amsmath,hyperref,graphicx} \usepackage{amssymb,float}
\usepackage{booktabs}

\usepackage[table]{xcolor}

\usepackage{a4wide}

\usepackage{booktabs} \usepackage{array,ragged2e}
\newcolumntype{R}[1]{>{\RaggedLeft\arraybackslash}p{#1}}
\usepackage{arydshln}

\usepackage[utf8]{inputenc}

\makeatletter
\def\ps@pprintTitle{%
  \let\@oddhead\@empty
  \let\@evenhead\@empty
  \let\@oddfoot\@empty
  \let\@evenfoot\@oddfoot
}
\makeatother

\usepackage{pdflscape}

\usepackage{fixme} \fxsetup{status=draft}

\usepackage{xcolor} \usepackage{natbib}

\usepackage[section]{placeins}

\usepackage{siunitx} 

\usepackage{float} \floatstyle{plaintop} \restylefloat{table}

\usepackage{textcomp}

\usepackage{caption} 

\usepackage[skip=0.333\baselineskip]{subcaption}

\author[aau,rga]{Mikkel Meyer Andersen\fnref{fn1}}
\author[rga]{Marie-Louise Kampmann}
\author[rga]{Alberte Honoré Jepsen}
\author[aau,rga]{Niels Morling}
\author[aau]{Poul Svante Eriksen}
\author[rga]{Claus Børsting}
\author[rga]{Jeppe Dyrberg Andersen}

\fntext[fn1]{Corresponding author, e-mail address: \url{mikl@math.aau.dk}}

\address[aau]{Department of Mathematical Sciences, Aalborg
University, Denmark} \address[rga]{Section of Forensic Genetics,
Department of Forensic Medicine, Faculty of Health and Medical
Sciences, University of Copenhagen, Denmark}

\usepackage{enumitem}
\setitemize{noitemsep,topsep=0pt,parsep=0pt,partopsep=0pt}

\begin{document}

\begin{frontmatter}




   \title{Shotgun DNA sequencing for human identification:
    Dynamic SNP selection and likelihood ratio calculations accounting for errors}


  \begin{abstract}
    In forensic genetics, short tandem repeats (STRs) are used as standard loci for human identification (HID) and identification of close relationships. However, degraded biological trace samples with low amounts of short DNA
    fragments (low-quality DNA samples) pose a challenge for STR typing. Predefined single nucleotide polymorphisms (SNPs)
    can be amplified on short PCR fragments and used to generate SNP
    profiles from low-quality DNA samples. However, the stochastic
    results from low-quality DNA samples may result in frequent locus
    drop-outs and insufficient numbers of SNP genotypes for convincing identification of individuals or relationships. 
    Shotgun DNA sequencing potentially analyses all DNA fragments in a sample in contrast to the targeted PCR-based
    sequencing methods and may be applied to DNA samples of very low
    quality, like heavily compromised crime-scene samples and ancient
    DNA samples. Low-quality
    samples have an increased probability of genotyping errors. 
    Here, we
    developed a statistical model for the process consisting of shotgun sequencing,
    sequence alignment, and genotype calling. Results from
    replicated shotgun sequencing of buccal swab 
    (high-quality samples) and hair samples 
    (low-quality samples) were arranged in
    a genotype-call confusion matrix to estimate the calling error
    probability by maximum likelihood and Bayesian inference. We
    developed formulas for calculating the evidential weight as a likelihood ratio
    ($LR$) based on data from dynamically selected SNPs from shotgun DNA sequencing. The method accounts for potential genotyping errors in a
    probabilistic manner. To our knowledge, the method is the first to
    make this possible.  Different genotype quality filters may be applied to account for genotyping errors. An error
    probability of zero resulted in the forensically commonly used
    $LR$ formula for the weight of evidence. When considering a single
    SNP marker's contribution to the $LR$, error probabilities larger
    than zero reduced the $LR$ contribution of matching genotypes and increased
    the $LR$ in the case of a mismatch between the trace genotype and the genotype of the person of interest. In the later scenario, the LR increased from zero, occurring with an error
    probability of zero, to positive values, as the mismatch may be
    due to genotyping errors. We developed an open-source R package,
    \texttt{wgsLR}, which implements the method, including estimating
    the calling error probability and calculating $LR$ values. The R
    package includes all formulas used in this paper and the functionalities to generate the formulas.
  \end{abstract}


\end{frontmatter}

%

\section{Introduction}

Degraded biological trace samples containing short DNA fragments and low
amounts of DNA (low-quality DNA samples) are challenging for standard forensic genetic analyses of short tandem repeats (STRs), where PCR amplicons of 85-400 bp are typically detected by capillary electrophoresis (CE) \citep{Gill2015}. Short DNA fragments with single
nucleotide polymorphism (SNP) loci may be amplified from degraded DNA and analysed by CE, e.g., with the SNP\textsl{for}ID assay that
analyses amplicons of 59-115 bp in length \citep{SNPforID,
  Boersting2013}. However, the stochastic results from low-quality DNA
samples may result in frequent locus drop-outs and insufficient
numbers of SNP genotypes for forensic genetic applications.

Shotgun DNA sequencing potentially analyses all DNA
fragments in a sample. This is in contrast to the targeted PCR
sequencing methods mentioned above, and shotgun sequencing may be applied to very
low-quality DNA from, e.g., heavily compromised crime-scene samples and ancient
DNA samples \citep{Rasmussen2010, Prfer2013}. Since shotgun DNA sequencing
does not rely on an initial PCR and a predefined set of SNPs, shotgun
DNA sequencing may be used for low-quality DNA
samples in a forensic genetic setting.

Low-quality samples have an increased probability of generating genotyping
errors. We modelled the process consisting of shotgun sequencing,
sequence alignment, and genotype calling, resulting in a statistical
model for genotyping errors with one parameter, $w$, which we refer to
as the calling error probability that can be attributed to one or more events in
the process (sequencing, alignment, and genotype calling). The calling error probability is important in evaluating
forensic DNA samples, e.g.\ with an evidential weight expressed as a
likelihood ratio ($LR$), because a genotype discrepancy between two
samples can be due to either SNP calling errors or different genotypes.

Results from replicated DNA sequencing of buccal swaps and hairs were
arranged in a genotype call confusion matrix to estimate the calling
error probability, $w$, using maximum likelihood estimation (MLE)
method and by Bayesian inference using the posterior mean. 

We developed formulas for the evidential weight as an $LR$ that uses a set of dynamically selected SNPs from shotgun DNA sequencing data. It is possible to incorporate factors such as genotype quality filtering using various criteria.

We assessed the strength of the method on the genomic areas around a subset of the SNP\textsl{for}ID markers \citep{SNPforID,
  Boersting2013}.

To our knowledge, this method is the first to enable $LR$ calculations based on shotgun DNA sequencing including accounting for genotyping errors with a probabilistic model.

\section{Materials and methods}

We used R \citep{R} version 4.3.2 for the analyses.

\subsection{Samples and ethical approval}

We analysed paired samples from three donors to
estimate the calling error probability, $w$. The paired samples
consisted of two different tissue types: a) buccal swab samples
($n = 6$ samples, i.e., paired samples from three individuals) and b) hair samples ($n = 16$ paired from two individuals), which represent high and low-quality samples, respectively.

The study was approved by the Committees of Health Research Ethics in the Capital Region of Denmark (H-22034131). Informed written consent was collected from all participants. The samples are stored in pseudonymised form in a biobank that is registered at the University of Copenhagen’s joint records of processing of personal data in research projects and biobanks (514–0789/23–3000) and complies with the rules of the General Data Protection Regulation (Regulation (EU) 2016/679). 

The high-quality samples were buccal swabs taken with a Foam Swab (Qiagen) that were subsequently rubbed against Whatman FTA\textregistered{} cards (Cytiva Life Science). One or four 3 mm punches from Whatman FTA\textregistered{} cards were extracted with the EZ1\&2 DNA Investigator Kit and the EZ1 Advanced XL (Qiagen) following the manufacturer’s instructions. The low-quality samples were either two cm of telogene hairs, including the root or the nearest two centimetres of hair without the root. The extraction was performed using the extraction described in \citep{Hellmann2001} following a purification using EZ1\&2 DNA Investigator Kit and the EZ1 Advanced XL following the manufacturer’s instructions. Sequencing libraries were made from all DNA extracts using the TruSeq Nano DNA Low Throughput Library (Illumina), and half of the hair sample pairs were converted into libraries using a non-commercial single-stranded  method \citep{Kapp2021}. The libraries were quantified using the KAPA Library Quanitification Kits - Complete kit (Roche) to prepare pooling in equimolar amounts. All samples were sequenced in two different sequencing runs on a NovaSeq 6000 system (Illumina) using NovaSeq 6000 S2 Reagent Kit v1.5, 300 cycles (Illumina).

\subsection{Independent chromosomal segments}
\label{sec:mm-markers}

For autosomal DNA profiles of STR and SNP loci,
calculating the evidential weight as an $LR$ is simple if the loci are independent. We obtained a set of independent SNPs by selecting one SNP in each of a limited number of independent
DNA segments to illustrate the model, estimate the
genotype calling error probability, and make computations more
feasible.


To evaluate the strength of the method, we selected chromosomal segments including 43 SNPs
from the SNP\textsl{for}ID marker set \citep{SNPforID}.  The 43 segments are
located at each end of the 22 autosomes, except for chromosome 19, where SNPs on only one chromosomal end are included. We
selected diallelic (also called biallelic) SNPs with minor allele frequencies (MAF) in Europeans closest to 0.5 as
segment centres. We included 100,000 bases before the centre and
99,999 bases after the centre, totalling 200,000 bp.
Thus, $43 \times$200,000 bp = 8,600,000 bp were
available for analysis for each sequenced genome.

We used the procedure described below to select SNPs in the segments. However, other procedures for selecting segments and SNP loci can be used and
combined with our statistical model and $LR$ formulas.

\begin{figure}[htbp]
  \centerline{\includegraphics[width=0.99\textwidth]{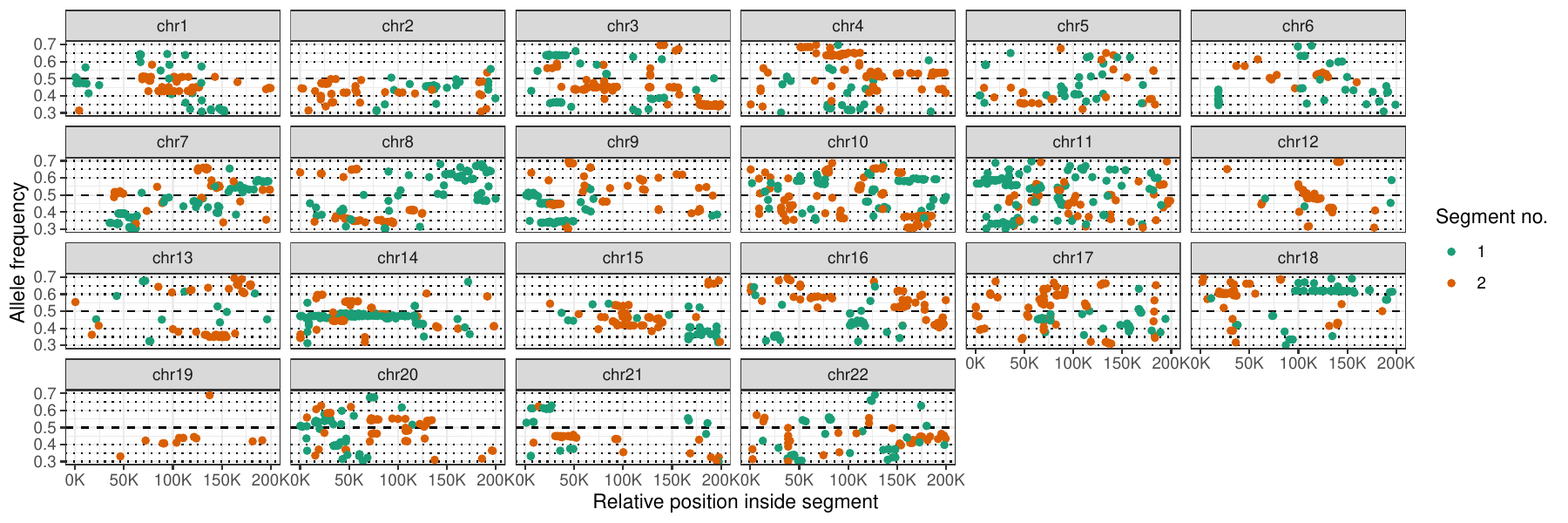}}
  \caption{Relative positions of candidate SNPs with minor allele
    frequencies (MAF) of at least 0.3. The dashed lines mark AF = 0.5, and the dotted lines mark AF distances (AFD) of 0.1, 0.15, and
    0.2, respectively.}
  \label{fig:gnomadsegments-count-relpos}
\end{figure}

We used gnomAD v3.1.2 \citep{gnomad} SNP information for African/African
American (AFR), East Asian (EAS), and European non-Finnish (NFE)
populations to illustrate a way to select loci amongst all the
possible ones in the segments. We only considered diallelic SNP
loci with an rs-number. We counted the number of candidate SNP
loci in each segment, considering the distance to an allele
frequency (AF) of 0.5 in the three populations (AFR,
EAS, and NFE). We considered three AF distances (AFD) to an
allele frequency of 0.5: 
AFD of 0.1 (MAF between 0.4 and 0.5), 
an AFD of 0.15 (MAF between 0.35 and 0.5), 
and AFD of 0.2 (MAF between 0.3 and 0.5). An AFD of 0.1 gave 473 candidates SNPs spread out among the 43 segments, an AFD of 0.15 gave 1,234
candidates SNPs, and an AFD of 0.2 gave 2,421 candidates SNPs.  We found that the candidate SNPs were distributed across the entire segments
(Fig.~\ref{fig:gnomadsegments-count-relpos}).  These candidate SNPs are available in the dataset \texttt{marker\_candidates\_v1} in our \texttt{R} package \texttt{wgsLR}.

Candidate SNPs must be checked for suitability and
selected based on the data from the crime-scene trace sample to avoid bias. It
is essential that the sample from the person of interest (PoI) is not considered before the trace sample, and comparisons between the trace and PoI samples are not
performed at this stage.

\subsection{Genotype calling}

All samples were shotgun DNA sequenced in-house using a NovaSeq 6000 system
(Illumina). Raw Binary Base Call (BCL) files from the sequencing were converted
into fastq files using \texttt{bcl2fastq2} (version 2.20.0.422,
Illumina) followed by adapter-trimming using \texttt{AdapterRemoval}
\citep{Schubert2016} (version 2.1.3) with a minimum trimming quality of
30, maximum N's allowed at 30, and a minimum fragment length of
30 \citep{Dabney2013}. The reads were aligned to the human reference genome GRCh38 using
BWA-MEM \citep{BWA} (version 0.7.10-r789) without removing duplicates
and collapsing the paired-end reads. Variants were detected
using the Genome Analysis Toolkit \citep{GATK} (GATK, version 4.0)
and Picard \citep{Picard2019toolkit} (version 2.25, Broad
Institute). 

It is important to note that the method works with different pipelines, including different software pipelines. It is necessary to obtain sequencing results, including
read depth and genotype quality, for all bases in the
selected segments. Thus, it is not sufficient to use information from
variants that -- with high probability -- differ from the reference genome (variants identified in vcf-file format), as this can introduce bias in the results. We
achieved this by using GATK HaplotypeCaller with the additional argument \texttt{--emit-ref-confidence BP\_RESOLUTION} for the segments (using \texttt{-L segments.interval\_list}).

\subsection{Quality criteria}

Data from non-diallelic SNPs were removed. We employed two quality
filters based on allele balance deviation, read depth (DP), and
genotype quality (GQ): a loose filter and a strict filter. The allele
balance (AB) was calculated as the proportion of reads for the allele with the fewest reads. The AB was between 0 and 0.5. The
filters were based on the AB deviation (ABD) from 0 for homozygous genotypes and
0.5 for heterozygous genotypes. Thus, for a homozygous genotype of
allele A with coverage 95 and allele C with coverage 5, the
allele balance is 5 / 100 = 0.05, and ABD = 0.05. Similarly, a
heterozygous genotype with minor allele coverage 45 and major allele
coverage 55 had an allele balance of 45 / 100 = 0.45 and an ABD
of 0.05 as 0.50-0.45 = 0.05.

The loose filter values were: ABD $\leq$ 0.3, DP $\geq$ 6, and GQ
$\geq$ 10. The strict filter values were: ABD $\leq$ 0.1, DP $\geq$
10, and GQ $\geq$ 20.

As noted above, the method works with different pipelines
and is flexible in employing other quality criteria.

\subsubsection{Software implementation}

We used the R computer algebra package \texttt{caracas}
\citep{caracasJOSS, caracasRJournal} to automate the formula
generation and verification and have included the formulas in the R
package \texttt{wgsLR}. Thus, \texttt{wgsLR} can generate the formulas
in the graphs and the tables. 

\subsection{Statistical model for genotyping errors}
\label{sec:mm-model}

We present a one-parameter statistical model for using
shotgun sequencing genotyping errors. We demonstrate how to estimate
the parameter and perform $LR$ calculations using the model. 
The functionality is available in the 
open-source R package \texttt{wgsLR}.

The trace sample is assumed to be a profile from a single
individual. First, we consider one locus. The contribution to the $LR$ from other loci can be
included in the $LR$ calculations by multiplication of their $LR$ if the loci are independent (not in linkage disequilibrium).

Below, we consider the profile of a trace donor and the profile of a PoI.  We only consider diallelic, non-phased SNPs. For notation, let $Z^S \in \{ 0, 1, 2\}$ be the genotype of
the PoI, e.g.\ suspect, $S$. It refers to the number of
alternative alleles, e.g.\ non-reference (e.g.\ GRCh38) alleles. Thus,
for $Z^S=0$, the PoI's genotype is homozygous (two
reference and zero alternative alleles), for $Z^S=1$ heterozygous (one
reference and one alternative allele), and for $Z^S=2$ homozygous
(zero reference and two alternative alleles). Sometimes, we also
employ the notation $Z^S \in \{ \text{0/0}, \text{0/1}, \text{1/1} \}$ to stress that there are two alleles, and 0, 1 or 2 are
the number of alternative alleles.  Similarly, let $Z^D \in \{ 0, 1, 2\}$ be the
genotype of the trace donor, $D$. 

Denote by $p_Z$ as the genotype
frequency in the population of interest.  If only allele frequencies and not genotype frequencies 
are available with $q$ being the frequency of the reference allele,
then under certain assumptions (Hardy-Weinberg equilibrium), including that the population
is not structured, 
\begin{align}
  p_0 = q^2 , \quad p_1 = 2q(1-q) \quad \text{and} \quad p_2 = (1-q)^2 .
\end{align}

Instead of observing $Z$ ($Z^S$ or $Z^D$), we observe $X$, which
is $Z$ with a potential error arising from sequencing, alignment, or
genotype calling. Because $Z$ is not actually observed, we refer to it 
as the latent genotype.

There are two genotypes, one from a PoI (e.g., a suspect), $Z^{S}$, and one from the donor, $Z^{D}$.  Both are typed
with shotgun sequencing yielding the profiles $X^{S}$ and $X^{D}$.
We are interested in the $Z$'s (and whether they are equal),
but we only observe $X$, which is $Z$ with a potential error.

Consider, e.g., the data presented in Table~\ref{tab:tabRefBuccalallA},
where we assume that $Z^{S} = Z^{D}$ as the data is from the same
individual (paired buccal swab samples). The diagonal is where the
two samples' genotype calls coincide, and the off-diagonal is where
the genotype calls differ for one or both alleles. For analysis of high-quality samples, we expect most of the table content
to be in the diagonal and very few in the 0/0,1/1 and 1/1,0/0
corners. We can think of the genotype error probability as a measure
of how many observations fall outside the diagonal.

\begin{table}
  
\def\estwBayesRefBuccalall{4.4 \times 10^{-5}}
\def\estwBayesInvRefBuccalall{22,718}
\def\tabsumRefBuccalall{8,596,131}
\def\tabsumpercRefBuccalall{99.96\%}
  
\def\estwBayesRefBuccalquallow{5.6 \times 10^{-6}}
\def\estwBayesInvRefBuccalquallow{179,221}
\def\tabsumRefBuccalquallow{7,550,311}
\def\tabsumpercRefBuccalquallow{87.79\%}
  
\def\estwBayesRefBuccalqualhigh{2.5 \times 10^{-6}}
\def\estwBayesInvRefBuccalqualhigh{398,042}
\def\tabsumRefBuccalqualhigh{4,983,634}
\def\tabsumpercRefBuccalqualhigh{57.95\%} \centering
  \scriptsize \subcaptionbox{All bases, no filter.\\$\hat{w} =
    \estwBayesRefBuccalall{} = 1 :
    $ \estwBayesInvRefBuccalall{}.\\Table sum = \tabsumRefBuccalall{}
    (\tabsumpercRefBuccalall{}).\label{tab:tabRefBuccalallA}}{
    \begin{tabular}{rrrr}
  \toprule
 & 0/0 & 0/1 & 1/1 \\ 
  \midrule
0/0 & 8,580,543 & 400 &  61 \\ 
  0/1 & 671 & 8,317 &  96 \\ 
  1/1 &  79 &  61 & 5,903 \\ 
   \bottomrule
\end{tabular}
 } \hspace{\fill}
  \subcaptionbox{Loose 
    filter.\\$\hat{w} = \estwBayesRefBuccalquallow{} = 1 : $
    \estwBayesInvRefBuccalquallow{}.\\Table sum =
    \tabsumRefBuccalquallow{}
    (\tabsumpercRefBuccalquallow{}).\label{tab:tabRefBuccalallB}}{
    \begin{tabular}{rrrr}
  \toprule
 & 0/0 & 0/1 & 1/1 \\ 
  \midrule
0/0 & 7,537,513 &  59 &   0 \\ 
  0/1 &  45 & 7,434 &  16 \\ 
  1/1 &   1 &  16 & 5,227 \\ 
   \bottomrule
\end{tabular}
 } \hspace{\fill}
  \subcaptionbox{Strict filter.\\$\hat{w} =
    \estwBayesRefBuccalqualhigh{} = 1 :
    $ \estwBayesInvRefBuccalqualhigh{}.\\Table sum =
    \tabsumRefBuccalqualhigh{}
    (\tabsumpercRefBuccalqualhigh{}).\label{tab:tabRefBuccalallC}}{
    \begin{tabular}{rrrr}
  \toprule
 & 0/0 & 0/1 & 1/1 \\ 
  \midrule
0/0 & 4,978,416 &   0 &   0 \\ 
  0/1 &   1 & 1,875 &   1 \\ 
  1/1 &   0 &   1 & 3,340 \\ 
   \bottomrule
\end{tabular}
 }
  \caption{Data from paired buccal swab samples from an individual. The estimates of
    $w$ are the Bayesian posterior means. The table sum is the number of
    markers. The percentages were calculated in relation to the total
    number of SNP sites, $43 \times$200,000 = 8,600,000.}
  \label{tab:tabRefBuccalall}
\end{table}

\begin{figure}[htbp]
  \centerline{\includegraphics[width=0.7\textwidth]{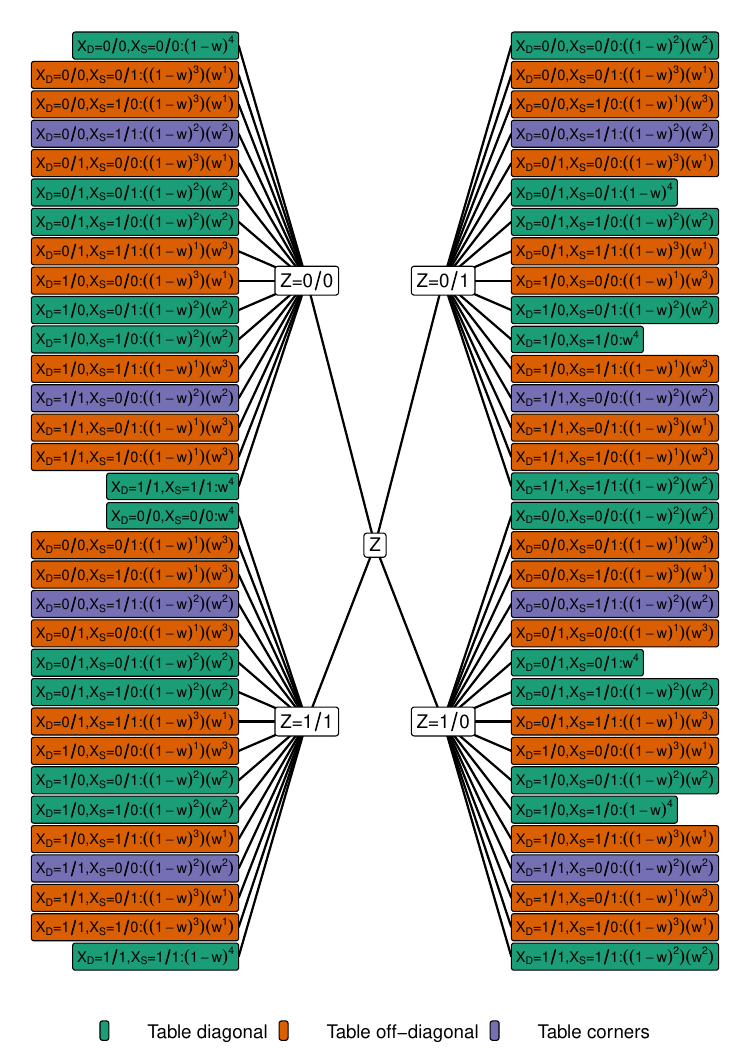}}
  \caption{Statistical model of two observed genotypes from a latent
    genotype, including the possibility for calling error with
    probability $w$, based on sequencing, alignment, and genotype
    calling. This graph can be found in the \texttt{wgsLR}
    object \texttt{graph\_Hp} and the formulas in the dataset
    \texttt{d\_formulas\_Hp}.}
  \label{fig:model-Hp}
\end{figure}

Let
$w$ be the calling error probability based on sequencing, alignment,
and genotype calling.  In the model it is assumed that the
genotype errors are independent, e.g., they do not cluster or depend
on the genomic content (e.g., C-streches).  When we assume that $Z =
Z^{S} =
Z^{D}$, the probability of the observed genotypes, accounting for errors, are shown in
Fig.~\ref{fig:model-Hp}. Note, that the distribution does not depend on
the value of $Z$. In other words, the prior probability of
$Z$ does not matter. The probabilities of observing a table like
Table~\ref{tab:tabRefBuccalallA} are shown in
Table~\ref{tab:model-Hp-table} (note that the probabilities sum to 1).

\begin{table}[h!]
  \footnotesize \centering 
\begin{tabular}[t]{ll}
\toprule
Group & Probability\\
\midrule
Table corners & $2 w^{2} \left(1 - w\right)^{2}$\\
Table diagonal & $w^{4} + 4 w^{2} \left(1 - w\right)^{2} + \left(1 - w\right)^{4}$\\
Table off-diagonal & $4 w^{3} \cdot \left(1 - w\right) + 4 w \left(1 - w\right)^{3}$\\
\bottomrule
\end{tabular}
  \caption{Probability distribution of table diagonal, corners, and
    off-diagonal (e.g., Table~\ref{tab:tabRefBuccalallA}). Note that
    the entries sum to 1. The distribution was obtained from
    Fig.~\ref{fig:model-Hp}, noting that the value of $Z$ does not
    matter and collecting the nodes that gave an observation in
    the diagonal, corners, or off-diagonal (not in diagonal and not
    in corners). The nodes' formulas can be found in the
    \texttt{wgsLR} object \texttt{d\_formulas\_Hp}.}
  \label{tab:model-Hp-table}
\end{table}

Based on this model, we can assume that the sum of the diagonal, the
sum of the corners, and the sum of the off-diagonal follow a trinomial
distribution with the parameters in
Table~\ref{tab:model-Hp-table}. Hence, the trinomial distribution does
not have two parameters as a standard trinomial distribution has, but
only the single calling error probability, $w$, based on which the
three trinomial parameters are calculated (Table~\ref{tab:model-Hp-table}).

\subsubsection{Parameter estimation}

The calling error probability, $w$, can be estimated using maximum
likelihood estimation and Bayesian inference. Both methods are implemented in
the open-source R package \texttt{wgsLR}, which we have developed. The
development version is available at
\url{https://github.com/mikldk/wgsLR} with online documentation at
\url{https://mikldk.github.io/wgsLR}, e.g., tutorials (vignettes)
demonstrating the functionality of the package.

\subsubsection{Evidential weight as a likelihood ratio}
\label{sec:mm-lr}

\begin{table}[h!]
  \footnotesize \centering \footnotesize 
\begin{tabular}[t]{rrl}
\toprule
$\mathbf{X^D}$ & $\mathbf{X^S}$ & $\mathbf{P(E \mid Z^S = Z^D)}$\\
\midrule
0 & 0 & $p_{0} \left(1 - w\right)^{4} + p_{1} w^{2} \left(1 - w\right)^{2} + p_{2} w^{4}$\\
0 & 1 & $2 p_{0} w \left(1 - w\right)^{3} + p_{1} w^{3} \cdot \left(1 - w\right) + p_{1} w \left(1 - w\right)^{3} + 2 p_{2} w^{3} \cdot \left(1 - w\right)$\\
0 & 2 & $p_{0} w^{2} \left(1 - w\right)^{2} + p_{1} w^{2} \left(1 - w\right)^{2} + p_{2} w^{2} \left(1 - w\right)^{2}$\\
1 & 0 & $2 p_{0} w \left(1 - w\right)^{3} + p_{1} w^{3} \cdot \left(1 - w\right) + p_{1} w \left(1 - w\right)^{3} + 2 p_{2} w^{3} \cdot \left(1 - w\right)$\\
1 & 1 & $4 p_{0} w^{2} \left(1 - w\right)^{2} + p_{1} w^{4} + 2 p_{1} w^{2} \left(1 - w\right)^{2} + p_{1} \left(1 - w\right)^{4} + 4 p_{2} w^{2} \left(1 - w\right)^{2}$\\
\addlinespace
1 & 2 & $2 p_{0} w^{3} \cdot \left(1 - w\right) + p_{1} w^{3} \cdot \left(1 - w\right) + p_{1} w \left(1 - w\right)^{3} + 2 p_{2} w \left(1 - w\right)^{3}$\\
2 & 0 & $p_{0} w^{2} \left(1 - w\right)^{2} + p_{1} w^{2} \left(1 - w\right)^{2} + p_{2} w^{2} \left(1 - w\right)^{2}$\\
2 & 1 & $2 p_{0} w^{3} \cdot \left(1 - w\right) + p_{1} w^{3} \cdot \left(1 - w\right) + p_{1} w \left(1 - w\right)^{3} + 2 p_{2} w \left(1 - w\right)^{3}$\\
2 & 2 & $p_{0} w^{4} + p_{1} w^{2} \left(1 - w\right)^{2} + p_{2} \left(1 - w\right)^{4}$\\
\bottomrule
\end{tabular}
  \caption{Probabilities of observing the evidence under the
    hypothesis $H_p$ of $Z^S = Z^D$. These formulas can be found in
    the \texttt{wgsLR} object \texttt{d\_prob\_Hp}.}
  \label{tab:model-LR-Hp}
\end{table}

We are in an identification setting and aim to use the $LR$ framework
with these hypotheses:

\begin{itemize}
\item $H_p$: The PoI, e.g.\ a suspect, $S$, with
  genotype $Z^S$, is the donor, $D$, of the crime-scene stain and
\item $H_d$: A random man, $D \neq S$, from the population of interest with genotype $Z^D$, is the donor of the crime-scene stain
\end{itemize}

If we assume no sequencing errors and there is a match between the PoI's profile and trace sample profile, we get the conventional 
\begin{align}
  LR 
  &= \frac{P(E \mid H_p)}{P(E \mid H_d)}  
    = \frac{P(X^S = X^D \mid H_p)}{P(X^S = X^D \mid H_d)} 
    = \frac{P(Z^S = Z^D \mid H_p)}{P(Z^S = Z^D \mid H_d)} 
    = \frac{1}{p_{Z^D}} ,
\end{align}
that is, the denominator is the population frequency of the
observed genotype.

Now, we consider the case with a calling error probability, $w$.
Under $H_p$, we need to find $P(E \mid H_p) = P(E \mid Z^S = Z^D)$, which are the probabilities when $Z^S = Z^D = Z$ and those are shown 
in Fig.~\ref{fig:model-Hp} 
(that was also used for derivation of parameter estimation). 
The probabilities in the nodes were 
multiplied with the corresponding $p_Z$ value in each node. By
doing this, and collecting the nodes corresponding to the observed
values of $X^S$ and $X^D$, we obtain the expressions in
Table~\ref{tab:model-LR-Hp}.

\begin{table}[h!]
  \centering \tiny 
\begin{tabular}[t]{rrl}
\toprule
$\mathbf{X^D}$ & $\mathbf{X^S}$ & $\mathbf{P(E \mid Z^S = Z^D)}$\\
\midrule
0 & 0 & $p_{0}^{2} \left(1 - w\right)^{4} + 2 p_{0} p_{1} w \left(1 - w\right)^{3} + 2 p_{0} p_{2} w^{2} \left(1 - w\right)^{2} + p_{1}^{2} w^{2} \left(1 - w\right)^{2} + 2 p_{1} p_{2} w^{3} \cdot \left(1 - w\right) + p_{2}^{2} w^{4}$\\
0 & 1 & $2 p_{0}^{2} w \left(1 - w\right)^{3} + 3 p_{0} p_{1} w^{2} \left(1 - w\right)^{2} + p_{0} p_{1} \left(1 - w\right)^{4} + 2 p_{0} p_{2} w^{3} \cdot \left(1 - w\right) + 2 p_{0} p_{2} w \left(1 - w\right)^{3} + p_{1}^{2} w^{3} \cdot \left(1 - w\right) + p_{1}^{2} w \left(1 - w\right)^{3} + p_{1} p_{2} w^{4} + 3 p_{1} p_{2} w^{2} \left(1 - w\right)^{2} + 2 p_{2}^{2} w^{3} \cdot \left(1 - w\right)$\\
0 & 2 & $p_{0}^{2} w^{2} \left(1 - w\right)^{2} + p_{0} p_{1} w^{3} \cdot \left(1 - w\right) + p_{0} p_{1} w \left(1 - w\right)^{3} + p_{0} p_{2} w^{4} + p_{0} p_{2} \left(1 - w\right)^{4} + p_{1}^{2} w^{2} \left(1 - w\right)^{2} + p_{1} p_{2} w^{3} \cdot \left(1 - w\right) + p_{1} p_{2} w \left(1 - w\right)^{3} + p_{2}^{2} w^{2} \left(1 - w\right)^{2}$\\
1 & 0 & $2 p_{0}^{2} w \left(1 - w\right)^{3} + 3 p_{0} p_{1} w^{2} \left(1 - w\right)^{2} + p_{0} p_{1} \left(1 - w\right)^{4} + 2 p_{0} p_{2} w^{3} \cdot \left(1 - w\right) + 2 p_{0} p_{2} w \left(1 - w\right)^{3} + p_{1}^{2} w^{3} \cdot \left(1 - w\right) + p_{1}^{2} w \left(1 - w\right)^{3} + p_{1} p_{2} w^{4} + 3 p_{1} p_{2} w^{2} \left(1 - w\right)^{2} + 2 p_{2}^{2} w^{3} \cdot \left(1 - w\right)$\\
1 & 1 & $4 p_{0}^{2} w^{2} \left(1 - w\right)^{2} + 4 p_{0} p_{1} w^{3} \cdot \left(1 - w\right) + 4 p_{0} p_{1} w \left(1 - w\right)^{3} + 8 p_{0} p_{2} w^{2} \left(1 - w\right)^{2} + p_{1}^{2} w^{4} + 2 p_{1}^{2} w^{2} \left(1 - w\right)^{2} + p_{1}^{2} \left(1 - w\right)^{4} + 4 p_{1} p_{2} w^{3} \cdot \left(1 - w\right) + 4 p_{1} p_{2} w \left(1 - w\right)^{3} + 4 p_{2}^{2} w^{2} \left(1 - w\right)^{2}$\\
\addlinespace
1 & 2 & $2 p_{0}^{2} w^{3} \cdot \left(1 - w\right) + p_{0} p_{1} w^{4} + 3 p_{0} p_{1} w^{2} \left(1 - w\right)^{2} + 2 p_{0} p_{2} w^{3} \cdot \left(1 - w\right) + 2 p_{0} p_{2} w \left(1 - w\right)^{3} + p_{1}^{2} w^{3} \cdot \left(1 - w\right) + p_{1}^{2} w \left(1 - w\right)^{3} + 3 p_{1} p_{2} w^{2} \left(1 - w\right)^{2} + p_{1} p_{2} \left(1 - w\right)^{4} + 2 p_{2}^{2} w \left(1 - w\right)^{3}$\\
2 & 0 & $p_{0}^{2} w^{2} \left(1 - w\right)^{2} + p_{0} p_{1} w^{3} \cdot \left(1 - w\right) + p_{0} p_{1} w \left(1 - w\right)^{3} + p_{0} p_{2} w^{4} + p_{0} p_{2} \left(1 - w\right)^{4} + p_{1}^{2} w^{2} \left(1 - w\right)^{2} + p_{1} p_{2} w^{3} \cdot \left(1 - w\right) + p_{1} p_{2} w \left(1 - w\right)^{3} + p_{2}^{2} w^{2} \left(1 - w\right)^{2}$\\
2 & 1 & $2 p_{0}^{2} w^{3} \cdot \left(1 - w\right) + p_{0} p_{1} w^{4} + 3 p_{0} p_{1} w^{2} \left(1 - w\right)^{2} + 2 p_{0} p_{2} w^{3} \cdot \left(1 - w\right) + 2 p_{0} p_{2} w \left(1 - w\right)^{3} + p_{1}^{2} w^{3} \cdot \left(1 - w\right) + p_{1}^{2} w \left(1 - w\right)^{3} + 3 p_{1} p_{2} w^{2} \left(1 - w\right)^{2} + p_{1} p_{2} \left(1 - w\right)^{4} + 2 p_{2}^{2} w \left(1 - w\right)^{3}$\\
2 & 2 & $p_{0}^{2} w^{4} + 2 p_{0} p_{1} w^{3} \cdot \left(1 - w\right) + 2 p_{0} p_{2} w^{2} \left(1 - w\right)^{2} + p_{1}^{2} w^{2} \left(1 - w\right)^{2} + 2 p_{1} p_{2} w \left(1 - w\right)^{3} + p_{2}^{2} \left(1 - w\right)^{4}$\\
\bottomrule
\end{tabular}
  \caption{Probabilities of observing the evidence under the
    hypothesis $H_d$. These formulas can be found in the
    \texttt{wgsLR} object \texttt{d\_prob\_Hd}.}
  \label{tab:model-LR-Hd}
\end{table}

\begin{figure}[h!]
  \vspace*{-4cm}
  \centerline{\includegraphics[height=1.1\textheight]{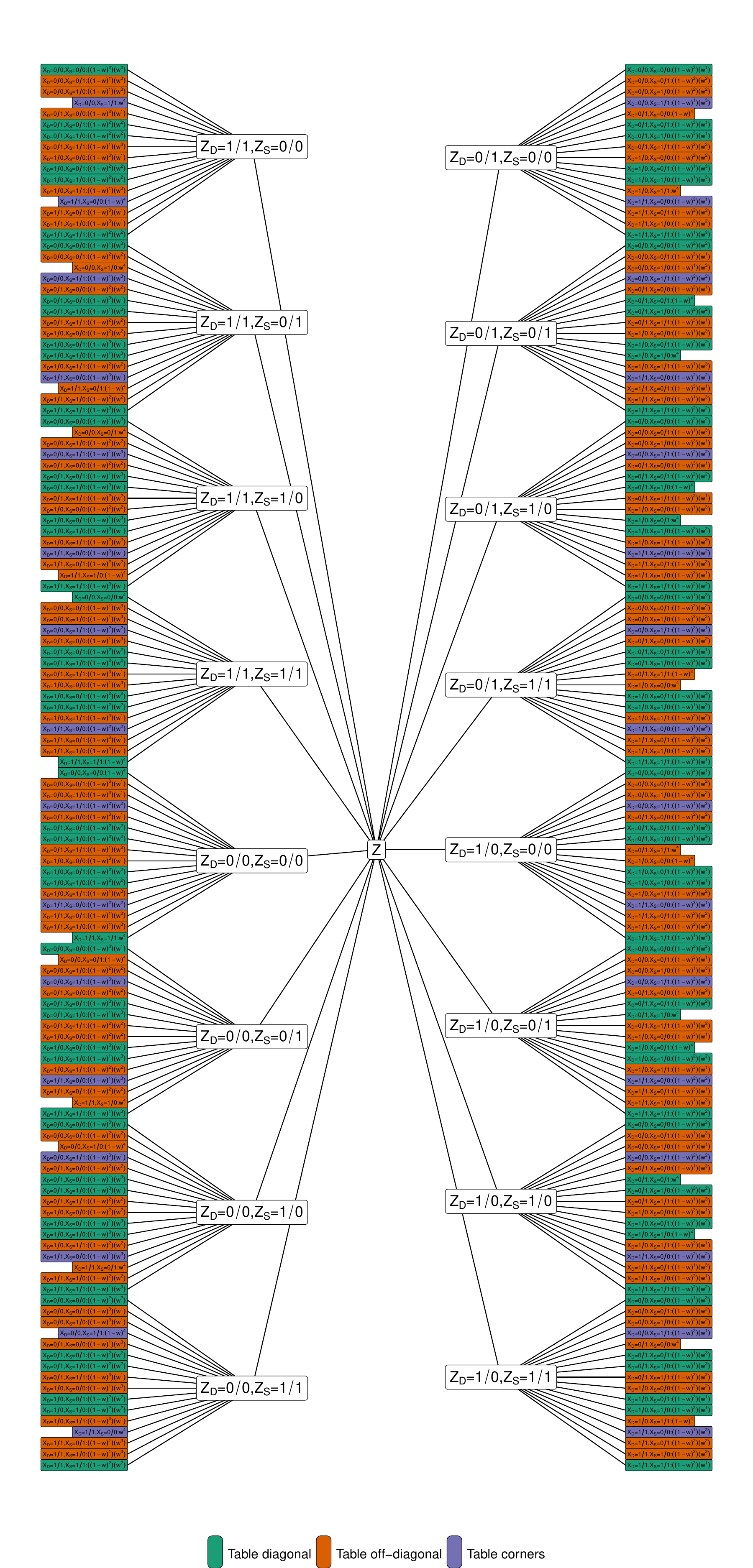}}
  \caption{Statistical model of two observed genotypes from two latent
    genotypes including calling error probability, $w$,
    based on sequencing, alignment, and genotype calling. This can be compared to Fig.~\ref{fig:model-Hp}, which shows 
    the situation for one latent genotype.}
  \label{fig:model-Hd}
\end{figure}

Under $H_d$, there are two latent genotypes, $Z^S$ and $Z^D$. A statistical model
similar to that in Fig.~\ref{fig:model-Hp} for $H_p$ can be seen in
Fig.~\ref{fig:model-Hd} for $H_d$. When we multiply the probabilities in the nodes by the
corresponding $p_Z$ value and collect the nodes corresponding to the
observed values of $X^S$ and $X^D$, we obtain the expressions in
Table~\ref{tab:model-LR-Hd}.

\begin{table}[h!]
  \footnotesize \centering 
\begin{tabular}[t]{rrl}
\toprule
$\mathbf{X^D}$ & $\mathbf{X^S}$ & $\mathbf{LR}$\\
\midrule
0 & 0 & $\frac{p_{0} \left(1 - w\right)^{4} + p_{1} w^{2} \left(1 - w\right)^{2} + p_{2} w^{4}}{p_{0}^{2} \left(1 - w\right)^{4} + 2 p_{0} p_{1} w \left(1 - w\right)^{3} + 2 p_{0} p_{2} w^{2} \left(1 - w\right)^{2} + p_{1}^{2} w^{2} \left(1 - w\right)^{2} + 2 p_{1} p_{2} w^{3} \cdot \left(1 - w\right) + p_{2}^{2} w^{4}}$\\
0 & 1 & $\frac{2 p_{0} w \left(1 - w\right)^{3} + p_{1} w^{3} \cdot \left(1 - w\right) + p_{1} w \left(1 - w\right)^{3} + 2 p_{2} w^{3} \cdot \left(1 - w\right)}{2 p_{0}^{2} w \left(1 - w\right)^{3} + 3 p_{0} p_{1} w^{2} \left(1 - w\right)^{2} + p_{0} p_{1} \left(1 - w\right)^{4} + 2 p_{0} p_{2} w^{3} \cdot \left(1 - w\right) + 2 p_{0} p_{2} w \left(1 - w\right)^{3} + p_{1}^{2} w^{3} \cdot \left(1 - w\right) + p_{1}^{2} w \left(1 - w\right)^{3} + p_{1} p_{2} w^{4} + 3 p_{1} p_{2} w^{2} \left(1 - w\right)^{2} + 2 p_{2}^{2} w^{3} \cdot \left(1 - w\right)}$\\
0 & 2 & $\frac{p_{0} w^{2} \left(1 - w\right)^{2} + p_{1} w^{2} \left(1 - w\right)^{2} + p_{2} w^{2} \left(1 - w\right)^{2}}{p_{0}^{2} w^{2} \left(1 - w\right)^{2} + p_{0} p_{1} w^{3} \cdot \left(1 - w\right) + p_{0} p_{1} w \left(1 - w\right)^{3} + p_{0} p_{2} w^{4} + p_{0} p_{2} \left(1 - w\right)^{4} + p_{1}^{2} w^{2} \left(1 - w\right)^{2} + p_{1} p_{2} w^{3} \cdot \left(1 - w\right) + p_{1} p_{2} w \left(1 - w\right)^{3} + p_{2}^{2} w^{2} \left(1 - w\right)^{2}}$\\
1 & 0 & $\frac{2 p_{0} w \left(1 - w\right)^{3} + p_{1} w^{3} \cdot \left(1 - w\right) + p_{1} w \left(1 - w\right)^{3} + 2 p_{2} w^{3} \cdot \left(1 - w\right)}{2 p_{0}^{2} w \left(1 - w\right)^{3} + 3 p_{0} p_{1} w^{2} \left(1 - w\right)^{2} + p_{0} p_{1} \left(1 - w\right)^{4} + 2 p_{0} p_{2} w^{3} \cdot \left(1 - w\right) + 2 p_{0} p_{2} w \left(1 - w\right)^{3} + p_{1}^{2} w^{3} \cdot \left(1 - w\right) + p_{1}^{2} w \left(1 - w\right)^{3} + p_{1} p_{2} w^{4} + 3 p_{1} p_{2} w^{2} \left(1 - w\right)^{2} + 2 p_{2}^{2} w^{3} \cdot \left(1 - w\right)}$\\
1 & 1 & $\frac{4 p_{0} w^{2} \left(1 - w\right)^{2} + p_{1} w^{4} + 2 p_{1} w^{2} \left(1 - w\right)^{2} + p_{1} \left(1 - w\right)^{4} + 4 p_{2} w^{2} \left(1 - w\right)^{2}}{4 p_{0}^{2} w^{2} \left(1 - w\right)^{2} + 4 p_{0} p_{1} w^{3} \cdot \left(1 - w\right) + 4 p_{0} p_{1} w \left(1 - w\right)^{3} + 8 p_{0} p_{2} w^{2} \left(1 - w\right)^{2} + p_{1}^{2} w^{4} + 2 p_{1}^{2} w^{2} \left(1 - w\right)^{2} + p_{1}^{2} \left(1 - w\right)^{4} + 4 p_{1} p_{2} w^{3} \cdot \left(1 - w\right) + 4 p_{1} p_{2} w \left(1 - w\right)^{3} + 4 p_{2}^{2} w^{2} \left(1 - w\right)^{2}}$\\
\addlinespace
1 & 2 & $\frac{2 p_{0} w^{3} \cdot \left(1 - w\right) + p_{1} w^{3} \cdot \left(1 - w\right) + p_{1} w \left(1 - w\right)^{3} + 2 p_{2} w \left(1 - w\right)^{3}}{2 p_{0}^{2} w^{3} \cdot \left(1 - w\right) + p_{0} p_{1} w^{4} + 3 p_{0} p_{1} w^{2} \left(1 - w\right)^{2} + 2 p_{0} p_{2} w^{3} \cdot \left(1 - w\right) + 2 p_{0} p_{2} w \left(1 - w\right)^{3} + p_{1}^{2} w^{3} \cdot \left(1 - w\right) + p_{1}^{2} w \left(1 - w\right)^{3} + 3 p_{1} p_{2} w^{2} \left(1 - w\right)^{2} + p_{1} p_{2} \left(1 - w\right)^{4} + 2 p_{2}^{2} w \left(1 - w\right)^{3}}$\\
2 & 0 & $\frac{p_{0} w^{2} \left(1 - w\right)^{2} + p_{1} w^{2} \left(1 - w\right)^{2} + p_{2} w^{2} \left(1 - w\right)^{2}}{p_{0}^{2} w^{2} \left(1 - w\right)^{2} + p_{0} p_{1} w^{3} \cdot \left(1 - w\right) + p_{0} p_{1} w \left(1 - w\right)^{3} + p_{0} p_{2} w^{4} + p_{0} p_{2} \left(1 - w\right)^{4} + p_{1}^{2} w^{2} \left(1 - w\right)^{2} + p_{1} p_{2} w^{3} \cdot \left(1 - w\right) + p_{1} p_{2} w \left(1 - w\right)^{3} + p_{2}^{2} w^{2} \left(1 - w\right)^{2}}$\\
2 & 1 & $\frac{2 p_{0} w^{3} \cdot \left(1 - w\right) + p_{1} w^{3} \cdot \left(1 - w\right) + p_{1} w \left(1 - w\right)^{3} + 2 p_{2} w \left(1 - w\right)^{3}}{2 p_{0}^{2} w^{3} \cdot \left(1 - w\right) + p_{0} p_{1} w^{4} + 3 p_{0} p_{1} w^{2} \left(1 - w\right)^{2} + 2 p_{0} p_{2} w^{3} \cdot \left(1 - w\right) + 2 p_{0} p_{2} w \left(1 - w\right)^{3} + p_{1}^{2} w^{3} \cdot \left(1 - w\right) + p_{1}^{2} w \left(1 - w\right)^{3} + 3 p_{1} p_{2} w^{2} \left(1 - w\right)^{2} + p_{1} p_{2} \left(1 - w\right)^{4} + 2 p_{2}^{2} w \left(1 - w\right)^{3}}$\\
2 & 2 & $\frac{p_{0} w^{4} + p_{1} w^{2} \left(1 - w\right)^{2} + p_{2} \left(1 - w\right)^{4}}{p_{0}^{2} w^{4} + 2 p_{0} p_{1} w^{3} \cdot \left(1 - w\right) + 2 p_{0} p_{2} w^{2} \left(1 - w\right)^{2} + p_{1}^{2} w^{2} \left(1 - w\right)^{2} + 2 p_{1} p_{2} w \left(1 - w\right)^{3} + p_{2}^{2} \left(1 - w\right)^{4}}$\\
\bottomrule
\end{tabular}
  \caption{Likelihood ratio ($LR$) of observing the evidence. These formulas
    can be found in the \texttt{wgsLR} object
    \texttt{d\_prob\_LR}. See the text on how to evaluate the formulas in
    R using \texttt{wgsLR} and \texttt{caracas} packages, e.g., in
    setting $w=0$ to see that the usual $LR$ value $\frac{1}{p_{Z^S}}$
    appears.}
  \label{tab:model-LR}
\end{table}

When we divide the corresponding observations from
Table~\ref{tab:model-LR-Hp} and Table~\ref{tab:model-LR-Hd}, the
evidential weight as an $LR$ is obtained. The formulas are
given in Table~\ref{tab:model-LR}.
The usual $LR$ values, $1 / p_{Z^S}$ for matches $X^D = X^S$ and $0$ for non-matches $X^D \neq X^S$, are obtained when setting $w = 0$. For non-matches $X^D \neq X^S$ and $w>0$, then $LR>0$.

\begin{figure}[htbp]
  \centerline{\includegraphics[width=0.8\textwidth]{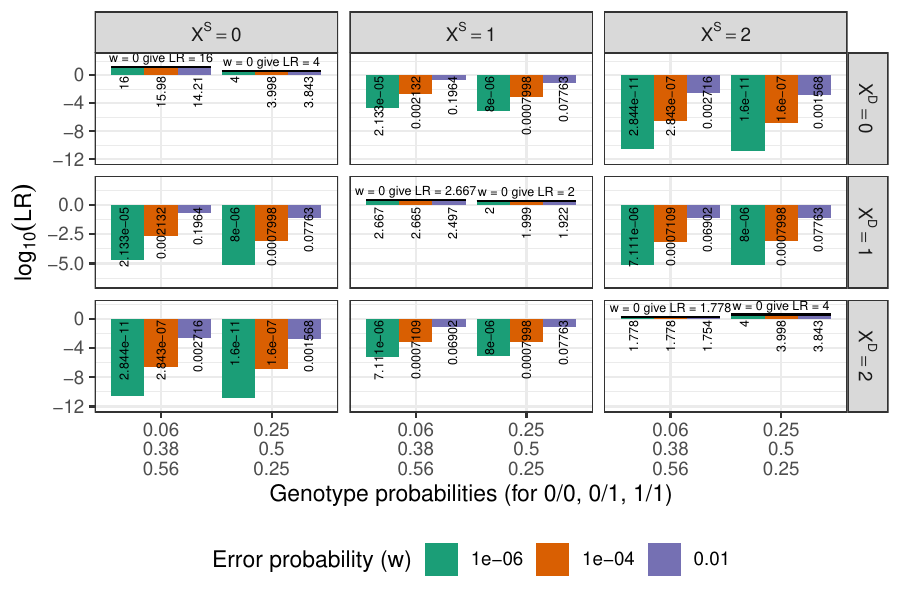}}
  \caption{Evidential weight as a likelihood ratio ($LR$) assuming Hardy-Weinberg equilibrium for allele frequency (AF) values $q = 0.25$ (genotype probabilities $p_0 = 0.06$, $p_1 = 0.38$, and $p_2 = 0.56$) and $q = 0.5$ (genotype probabilities $p_0 = 0.25$, $p_1 = 0.5$, and $p_2 = 0.25$) and three values of $w$ (calling error
    probability). Note, if we assume that no error is possible,
    $X^D \neq X^S$ results in $LR = 0$.}
  \label{fig:model-LR-numbers}
\end{figure}

To illustrate our model numerically, we performed $LR$ calculations
for a single locus with reference allele frequencies of 0.25 and 0.5,
assuming no population structure (Hardy-Weinberg equilibrium), and
three calling error probabilities of $10^{-6}$, $10^{-4}$, and
$10^{-2}$, respectively. The results are shown in
Fig.~\ref{fig:model-LR-numbers}.

The above method was implemented in the \texttt{calc\_LRs()} function
in our \texttt{wgsLR} R package.

\section{Results}

Selected chromosomal segments around a subset of the SNP\textsl{for}ID markers \citep{SNPforID, Boersting2013} for 11 pairs of samples (three pairs from buccal swaps and eight pairs from hairs) were used to assess the capabilities of the model in estimating the genotyping error probability. We used high-quality samples (buccal swap) and low-quality samples (hair) to illustrate different error probabilities. For assessing the 
capabilities of the model in calculating the evidential weight when accounting for potential sequencing errors, we conducted simulation experiments.


\begin{figure}[htbp]
  \centerline{\includegraphics[width=0.6\textwidth]{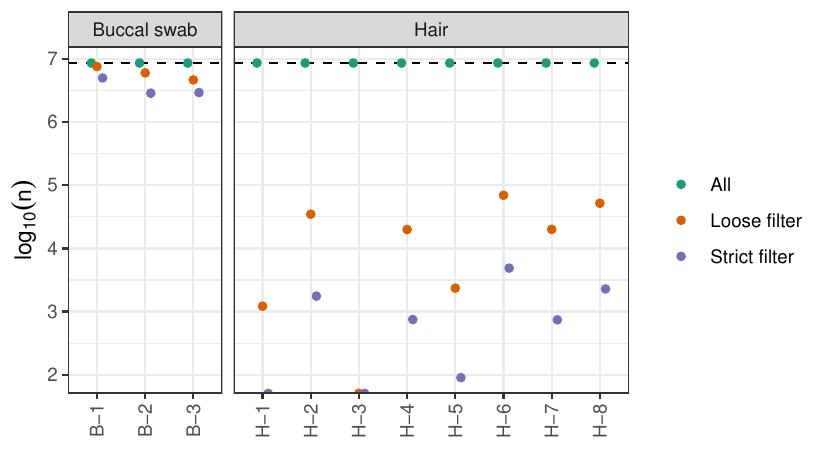}}
  \caption{Number of sites in the tables, $n$, for paired samples with
    no filter, loose filter, and strict filter. The dashed line shows
    $\log_{10}$ of 8,600,000, the maximal number of sites in
    the tables. The points at the bottom for hair (H-1) and (H-3) indicate
    tables of size 0 (no positions obtained).}
  \label{fig:paired-table-sizes}
\end{figure}

First, we considered the number of available sites available for analysis. 
The maximal number of possible sites with results, was $43 \times$200,000 =
8,600,000. The number of sites with results in all of the paired samples
(buccal swab and hair) ranged between 8,597,154 and 8,599,957, i.e., a maximum of 2,846 sites were removed from analyses when no quality
filter was applied. Fewer genotypes were available when applying
quality filters, especially for hair samples
(Fig.~\ref{fig:paired-table-sizes}).

\subsection{Maximum likelihood estimation or Bayesian inference}

If a table only has non-zero entries in the diagonal, then
$\hat{w}_{\text{MLE}} = 0$ (Table~\ref{tab:model-Hp-table}). This can happen, e.g., if the table
is small relative to $w$. Bayesian estimation includes information from
the prior and will provide an estimate $\hat{w}_{\text{Bayesian}} >0$ for many priors, including the uniform prior on $(0, 0.5)$ ($w$ and $1-w$ occur symmetrically in the model, so we used this restriction as $w$ was interpreted as the error probability and not the success probability). On the other hand, the
Bayesian estimates can become too large if the uniform prior is used on small table sizes, so care must be taken with tables with
few observations.  We used the uniform prior on $(0, 0.5)$ and the Bayesian posterior mean in the
rest of the analyses to circumvent this problem of obtaining
$\hat{w}_{\text{MLE}} = 0$.  If the tables are of sufficient size
relative to $w$, both the maximum likelihood estimation and the
Bayesian posterior mean will provide comparable estimates.

\begin{figure}[htbp]
  \centerline{\includegraphics[width=0.7\textwidth]{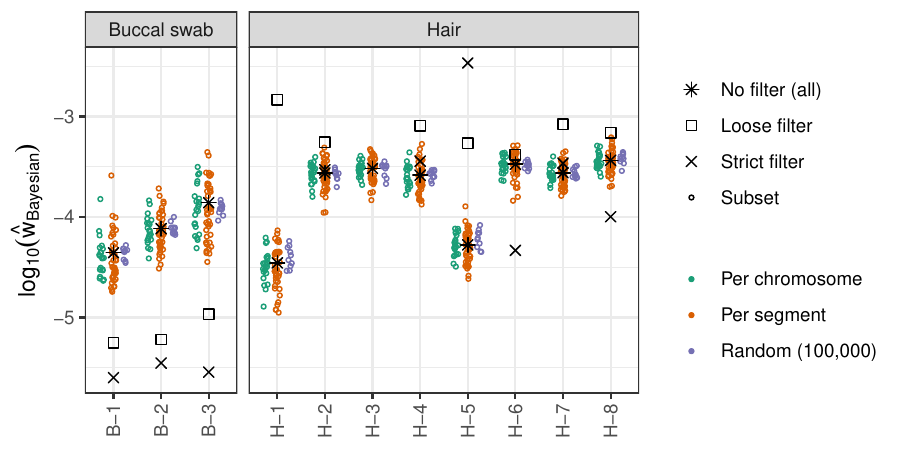}}
  \caption{Estimates of the probability of genotyping errors using the
    Bayesian posterior mean. Note, that hair 5 (H-5) had a high estimated
    probability of genotyping error, as only 90 genotypes were
    observed in the 0/0 entry and none elsewhere. Thus, the posterior
    mean is strongly influenced by the uniform prior. Hair
    1 (H-1) and 3 (H-3) miss estimates for one or more of the filters as the corresponding tables did not contain any genotypes (Fig.~\ref{fig:paired-table-sizes}).}
  \label{fig:paired-w-bayesian}
\end{figure}

\begin{figure}[htbp]
  \centerline{\includegraphics[width=0.9\textwidth]{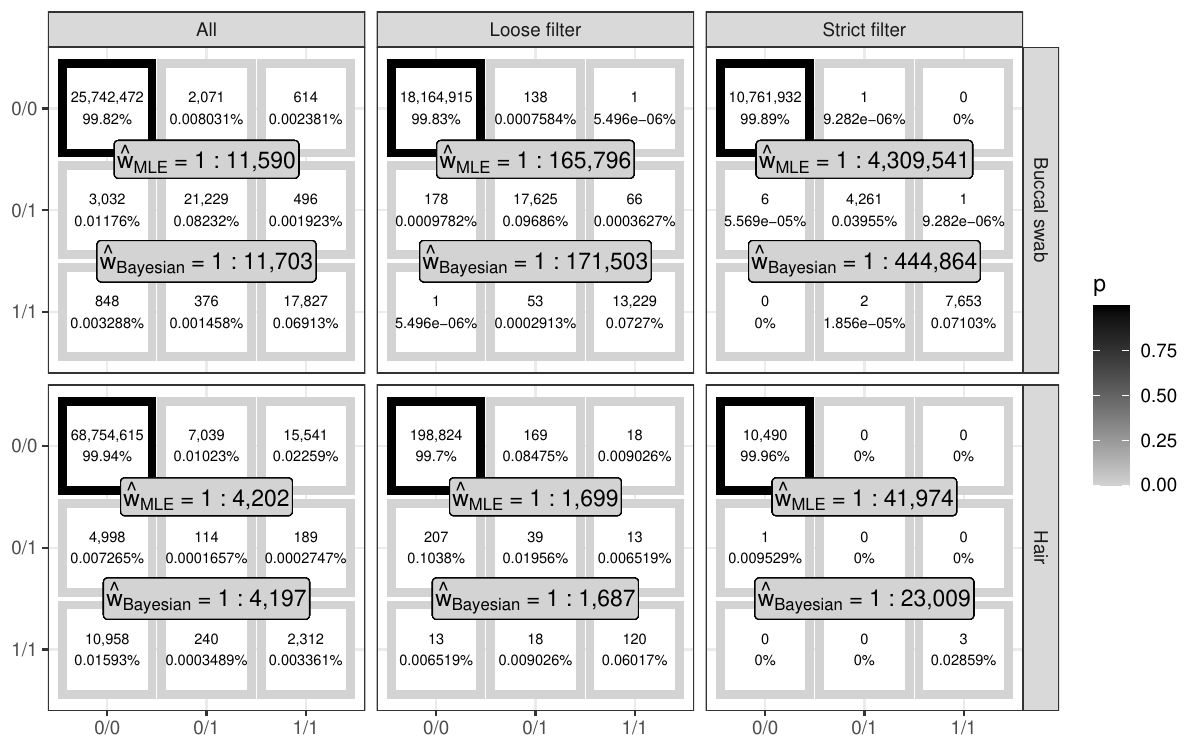}}
  \caption{Estimates of the genotype call probabilities using maximum likelihood estimation (MLE) and
    the Bayesian posterior mean based on aggregated tables from the
    two tissue types. The percentages sum to 100\% within each table.}
  \label{fig:paired-w-aggregated}
\end{figure}

Fig.~\ref{fig:paired-w-bayesian} shows the results from the paired
samples. Data from all samples of the same type (buccal swab and hair)
filtered in the same way were collected in a single table by summing
the cell contents cell-wise. Thus, $m \times 43 \times$200,000 observations were possible in each table,
where $m$ is the number of samples. The results of the tables are
shown in Fig.~\ref{fig:paired-w-aggregated}.

\subsection{Evidential weight as a likelihood ratio}
\label{sec:results-lr}

\subsubsection{Simulation study}

\begin{figure}[htbp]
  \centerline{\includegraphics[width=0.7\textwidth]{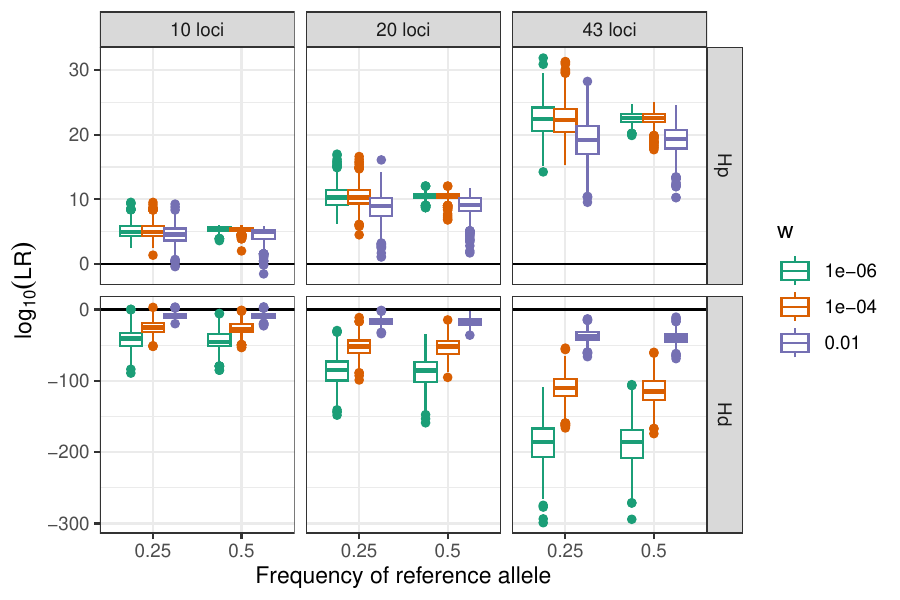}}
  \caption{Boxplot of likelihood ratio ($LR$) values for simulated cases under $H_p$ and
    $H_d$ for different combinations of calling error probabilities,
    reference allele frequencies, and number of loci.}
  \label{fig:LR-simulation-correct-w}
\end{figure}

To illustrate our model numerically for multiple SNP loci, we
performed $LR$ calculations on simulated and real data. We simulated
data with 10, 20, and 43 loci, with reference allele frequencies of 0.25 and 0.5, respectively, assuming no
population structure (Hardy-Weinberg equilibrium), and three different call
error probabilities of $10^{-6}$, $10^{-4}$, and $10^{-2}$,
respectively. 
The possible $LR$s for a single locus can be found in Fig.~\ref{fig:model-LR-numbers}. 
We simulated 1,000 cases under $H_p$ and 1,000 cases
under $H_d$ for all combinations of numbers of loci, true calling
error probabilities, and reference allele frequencies. The results are
shown in Fig.~\ref{fig:LR-simulation-correct-w}.

\subsubsection{Comparison results of samples from buccal swab and hair}
\label{sec:results-markers}

We used the candidate SNP sets obtained from gnomAD
\citep{gnomad} data as described in Sec.~\ref{sec:mm-markers} to find
the number of loci available in pairwise comparisons
between all paired samples such that each sample was, in turn, considered a
donor of a crime-scene trace and all other samples were considered
PoI samples. There were 22
samples,
leading to $22 \times 21 = 462$ pairwise
comparisons. Each pairwise comparison was done with three filters
(none, loose, and strict) and three AFD to 0.5. That gives $462 \times 3 \times 3$ =
4,158 ``cases''. 

The trace donor's SNP profile was analysed after applying no filter,
loose filter, or strict filter, and for each segment, selecting the
marker with the highest rank. The ranking was done according to the
score defined as the unweighted sum of the absolute deviation from 0.5
for AFR, EAS, and NFE. Genotypes identified in the profiles from both the donor and the PoI were analysed.

\begin{figure}[htbp]
  \centerline{\includegraphics[width=0.95\textwidth]{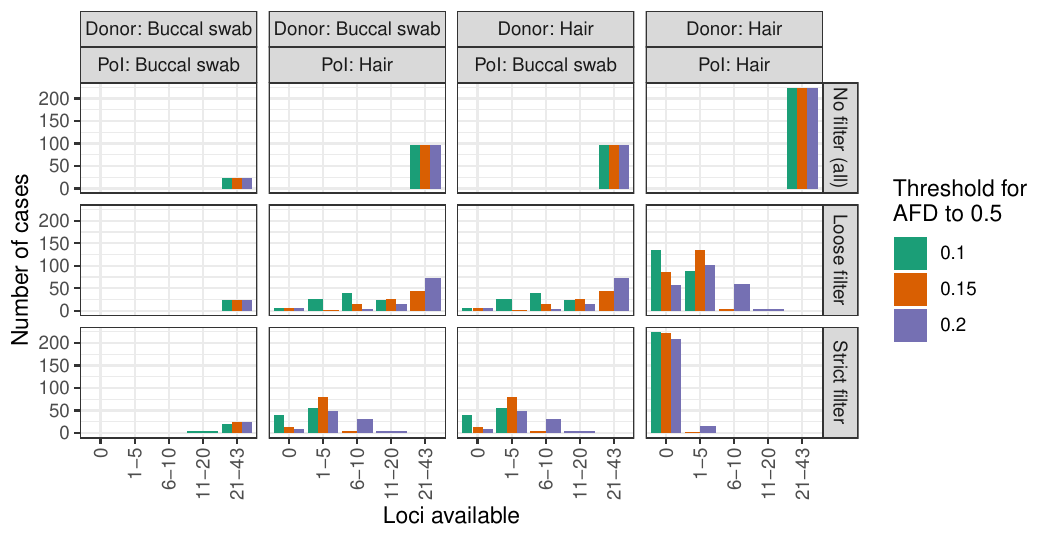}}
  \caption{Number of cases with a given number of loci available for cases under $H_d$. AF: Allele frequency, AFD: AF distance.}
  \label{fig:reallife-LR-comparison-num-markers-Hd}
\end{figure}

Fig.~\ref{fig:reallife-LR-comparison-num-markers-Hd} shows that SNPs were identified in all segments in the buccal swab samples
even with the strict quality filter, whereas very few SNP genotypes were identified in the hair samples unless no filter was applied.

\section{Discussion}

We have introduced a model for calculating the evidential weight
of shotgun DNA sequencing profiles by an $LR$ that
accounts for errors in the sequencing process, software pipeline, and
genotype calling (Sec.~\ref{sec:mm-model} and
Sec.~\ref{sec:results-lr}). 

We demonstrated how to estimate the
probability of the genotyping error for two tissue types of varying quality using maximum likelihood estimation and Bayesian inference
(Sec.~\ref{sec:mm-lr}) and showed that the probability of
genotyping error, particularly in low-DNA-quality samples, depends on the stringency of the quality filters
(Fig.~\ref{fig:paired-w-bayesian} and
Fig.~\ref{fig:paired-w-aggregated}).

The evidential weight calculations using the model was demonstrated on simulated data (Fig.~\ref{fig:LR-simulation-correct-w}).

The method (estimation the probability of the genotyping error and $LR$ calculations) was implemented in the open-source R package
\texttt{wgsLR} available at \url{https://github.com/mikldk/wgsLR} with
online documentation at \url{https://mikldk.github.io/wgsLR} and tutorials (vignettes) demonstrating the functionality of the
package.

We demonstrated the value of the method with SNP loci in predefined
segments centred around the SNP\textsl{for}ID SNPs \citep{SNPforID}. We
used all SNPs in the segments to enable estimation of the genotype error
probability. 
SNP selection does not need to be based on predefined
segments, followed by identification of SNP loci in the segments. Instead, SNPs can be dynamically selected based on, e.g., coverage and genotype quality maps from shotgun sequencing data of a trace sample, using information about quality, genotype frequencies, and linkage
disequilibrium. This would be a helpful approach for casework samples. Statistically independent loci are required to
multiply $LR$s from the loci without adjustment for linkage
disequilibrium. In such cases, it may be difficult to establish the
correct genotyping error probability without duplicate typing of the trace sample. We therefore suggest that samples should be typed in duplicates to get the sample specific genotype error probabilities. However, if trace material is scarce and does not allow for duplicate sample typing, estimates from samples
of the same tissue type and quality may serve as an approximation of $w$. It would also be
possible to integrate out the genotyping error probability --
similar to what can be done in other models, e.g.,
\citep{Slooten2017}, but further research is needed. Our model currently only supports one genotyping error probability, so further research is required on trace- and
PoI-specific genotyping error probabilities.

In our analyses using segments, we used data from the gnomAD
\citep{gnomad} database to select SNP loci with minor allele frequencies close to 0.5
(Sec.~\ref{sec:mm-markers} and
Sec.~\ref{sec:results-markers}). A similar approach can be used with
loci selected by other means, e.g., coverage maps.

More work needs be be done to validate potential markers
(e.g., if non-coding markers are required) and potentially ordering
the candidate markers within each segment. Our method is flexible so that other strategies
for selecting markers (e.g., SNP loci) and chromosomal segments can be used.

We emphasise that selecting the loci used for the
$LR$ calculations based on the data from the trace sample is essential. So, first, the trace
sample must be analysed and the loci selected among the ones with genotypes called in the trace and based on predefined
criteria, e.g., quality and population frequencies. Second, the $LR$ can be
calculated based on data from the trace sample and the PoI. 

Our model currently only supports binary markers like most SNPs, indels, etc., and more
research is needed to support more polymorphic markers such as microhaplotypes, tri-allelic SNPs, etc., that may be useful as a supplement to binary SNP
loci.

The described method is currently only for single-source
stains. Traces with mixtures of DNA are of high interest and are an area for more research. Similar models may be used to analyse shotgun DNA
sequencing data from two or more individuals.

The fact that we obtain
a high number of data from the segments (potentially up to 8,600,000 positions
with the proposed segments) but only consider a small subset of loci (e.g.\
43 positions) for $LR$ calculations may lead to questions about bias and whether this can be problematic for the PoI. We do not believe that there is a bias problem. However, the loci must be selected based on the trace samples' sequencing data, fulfilling the quality
criteria. The loci must not be selected based on knowledge of the
profile of the PoI. The $LR$ estimate can be further qualified by selecting several candidate SNPs in each
segment and calculate the $LR$ for several or all combinations of SNPs in the selected segments. If, e.g., three SNPs in each of the 43 SNP\textsl{for}ID segments are selected, $3^{43} > 3 \times 10^{20}$ SNP profiles can be constructed and the corresponding $LR$s calculated. In a simple scenario, the lowest $LR$ may be reported.

Additional analyses may be performed, like
inspecting the genotypes around the SNP markers chosen for $LR$
calculations for genetic inconsistency between the trace and PoI samples. However, the probability of genotyping errors must be considered, as done in our model for $LR$ calculations.

In conclusion, we have developed a probabilistic model based on DNA
sequencing error rates for evaluating the weight of evidence of
shotgun DNA sequencing results. The model is helpful in criminal cases
with highly compromised DNA from heavily degraded crime-scene samples,
disaster victims, etc. The method is implemented in an open-source R
package, \texttt{wgsLR}, which is freely available.

\bibliographystyle{unsrt} \bibliography{manus-r00-01}

\end{document}